\documentclass[preprint, 12pt,
showpacs, eqsecnum
,amsmath,amssymb
]{revtex4-1}
\usepackage{amssymb}
\usepackage{amsmath}
\usepackage{graphicx}
\usepackage{bm}
\usepackage{amsfonts}

\usepackage[caption=false]{subfig}

\usepackage[applemac]{inputenc}

\usepackage{libertine}
\usepackage[T1]{fontenc}

\newcommand{\Pe}{\mathbb{P}}
\newcommand{\A}{\mathbb{A}} 

\newcommand{\C}{\mathbb{C}}

\newcommand{\AT}{\mathbb{A}^{\footnotesize{\hbox{\scriptsize{T}}}}}

\newcommand{\hAT}{\hat \A^{\footnotesize{\hbox{\scriptsize{T}}}}}

\newcommand{\F}{\mathbb{F}}

\newcommand{\V}{\mathbb{V}}

\newcommand{\Lag}{\mathcal{L}}
\newcommand{\E}{\mathbb{E}}

\newcommand{\Tr}{\text{Tr}}
\newcommand{\be}{\begin{equation}} \newcommand{\ee}{\end{equation}} \newcommand{\ben}{\begin{eqnarray}}
\newcommand{\een}{\end{eqnarray}} \newcommand{\p}{\partial} 
\newcommand{\ep}{\varepsilon}\newcommand{\lb}{\label}
\newcommand{\nn}{\nonumber}

\begin{document}

\title{Instantons in a Lagrangian model of turbulence}
\author{L S Grigorio$^{1,2}$, F Bouchet $^{1}$, R M Pereira$^{1,3}$ and L Chevillard $^1$ }
\affiliation{$^1$Univ. Lyon, \'{E}cole Normale Sup\'{e}rieure de Lyon, Univ. Claude Bernard, CNRS, Laboratoire de Physique, F-69342 Lyon, France,}
\affiliation{$^2$Centro Federal de Educa\c c\~ao Tecnol\'ogica Celso Suckow da Fonseca, Av. Governador Roberto Silveira 1900, Nova Friburgo, 28635-000, Brazil}
\address{$^3$ CAPES Foundation, Ministry of Education of Brazil, Bras\'ilia/DF 70040-020, Brazil}

\begin{abstract}

The role of instantons is investigated in the Lagrangian model for the velocity gradient evolution known as the Recent Fluid Deformation (RFD) approximation. After recasting the model into the path-integral formalism, the probability distribution function (pdf) is computed along with the most probable path in the weak noise limit through the saddle-point approximation.  Evaluation of the instanton solution is implemented numerically by means of the iteratively Chernykh-Stepanov method. In the case of the longitudinal velocity gradient statistics, due to symmetry reasons, the number of degrees of freedom can be reduced to one, allowing the pdf to be evaluated  analytically as well, thereby enabling a prediction of the scaling of the moments as a function of Reynolds number. It is also shown that the instanton solution lies in the Vieillefosse line concerning the $RQ$-plane. We illustrate how instantons can be unveiled in the stochastic dynamics performing a conditional statistics.
\end{abstract}

\maketitle
\section{Introduction} 
This paper aims at obtaining the stationary probability distribution function and large fluctuations of a stochastic model of turbulence proposed by Chevillard and Meneveau \cite{chevi_menev}.
 The model, known as Recent Fluid Deformation (RFD) approximation, consists in a set of stochastic differential equations describing the evolution of the eight degrees of freedom of the velocity gradient tensor of a fluid particle along its Lagrangian trajectory in an incompressible flow. Large deviations of the velocity gradient in turbulent flows are associated with high dissipation rates and enstrophy and are crucial to the understanding of intermittency phenomena - a topic of intense research in turbulence. In order to evaluate the pdfs of the velocity gradients (and also the probability of large fluctuations) we made use of the iterative numerical procedure of Chernykh-Stepanov \cite{stepanov}  which amounts to solving the saddle-point equations that minimize the action, providing this way the most probable path leading to a given fluctuation, which will be refered to as the instanton.  In the case of the longitudinal velocity gradient, due to symmetry reasons, the number of degrees of freedom can be reduced to one, allowing the pdf to be obtained analytically as well. These analytical probability distribution functions (pdfs) obtained are in excellent agreement with the numerical ones obtained by numerical integration of the stochastic differential equations. Another result is that the instanton lies along the Vieillefosse line in the so-called $RQ$-plane. For the longitudinal velocity gradient, this instanton approach gives unprecedented prediction for the pdf tail and for the dynamics of the optimal path, along with a prediction of how the moments scale with the Reynolds number. That gives us a theoretical approach to the dynamics leading to these rare events, and thus intermittency. This point is the main originality of this work.

Of central interest in turbulence is the behavior of small scales statistics. More specifically,  scaling and universality at small scales of motion in turbulent flows is a long standing problem \cite{antonia_sreenivasan}. 
Due to the intense fluctuations within small scales, large deviations of the velocity field differences are very pronounced for high values of Reynolds number.  These large excursions of the velocity gradient are apparent in the pdf, where drastic departures from gaussian behavior are manifest - and also termed intermittency. 

It is clear that a theory of turbulence capable of explaining intermittency and the scaling of high order structure functions must rely on a deep understanding of the dynamics of the small scales. 
A natural candidate to probe such scales is the velocity gradient tensor.  Nevertheless, obtaining the statistics of the velocity gradient tensor is a difficult task.  A common approach to address the evolution of the velocity gradient is the Lagrangian framework, which can drastically reduce the degrees of freedom and lead to a simplified picture of the small scales.  Turbulence in the Lagrangian frame has some different features compared to Eulerian turbulence, such as a shorter correlation time of the velocity gradient.  This property inspired the Recent Fluid Deformation \cite{chevi_menev} approximation, which is a model where the shape of a fluid particle following the local velocity field has a short memory. This closure was studied  in the last years \cite{afonso_menev}, \cite{chevi_menev2}, \cite{chevi_etal} and extended to account for passive scalar transport and MHD \cite{grauer_rfd} and was dealt with analytically by an effective action approach, based on noise renormalisation \cite{effectiveaction}.  

In order to study large deviations of the velocity gradients in this model  the path integral framework is used, which is very suitable to investigate large fluctuations. The reason lies in the fact that for weak noise driven systems, the probability is dominated by the action minima. The trajectory which minimizes the action is called instanton. This approach is equivalent to the Freidlin-Wentzell theory of large fluctuations \cite{freidlin_wentzel} and provides a proper way to find which is the most probable evolution leading to a large event. It can be, therefore, a valuable approach to deal with an important question in hydrodynamic turbulence, that is what are the common structures found at small scale turbulence.  
Structures related to large values of velocity gradient are of vital importance in the study of turbulence, since they are responsible for most of dissipation that takes place at the smallest scales of fluid motion, typically associated with large strain and vorticity.  Many works have devoted a long effort on the identification of such objects. In particular the use of instanton techniques to achieve this goal in Burgers turbulence can be found in references \cite{stepanov}, \cite{migdal}, \cite{grafke_schafer}, \cite{grafke_eric}. Reference \cite{effectiveaction} applies the path-integral approach  to evaluate the pdf in the RFD model. However, the set of saddle-point equations was linearized to obtain an approximate instanton solution. To correct this truncated saddle-point equations, a perturbative method was carried out. 


In this work we determine the instanton of the RFD  addressing what is the most probable evolution of a Lagrangian particle and also calculate its contribution to the pdf in the weak noise limit by solving the full set of non-linear saddle-point equations.
For the case of a diagonal (longitudinal) component of the velocity gradient, analytical results can be computed for its pdf.

The paper is organized as follows. In section II the Recent Fluid Deformation equations are reviewed. Section III is devoted to the results and is divided in four parts. Part A presents the model  dressed in the Martin-Siggia-Rose/Janssen/de Dominicis path integral formalism \cite{msr}, \cite{janssen}, \cite{dominicis} and how this approach can be used to address large deviations.  Part B displays the  transverse velocity gradient statistics after solving the instanton equations by means of the numerical Chernykh-Stepanov \cite{stepanov} algorithm. In part C it is shown that the longitudinal velocity gradient is subject to an analytical solution in addition to the numerical one.  In the sequel, part D presents how instantons are uncovered by performing a conditioned statistics with respect to the stochastic dynamics, which are confronted with the previously obtained instantons.  Final remarks close the paper in section IV.

\section{The RFD Lagrangian stochastic model}

\subsection{Recent Fluid Deformation for Lagrangian turbulence}
Proposed in \cite{chevi_menev}, the Recent Fluid Deformation (RFD) is a scheme for modelling the evolution of velocity gradient of a fluid particle along its trajectory in the Lagrangian frame. By taking the gradient of the Navier-Stokes equation, we write
\be
\lb{1}
\frac{d A_{ij}}{dt} = -A_{ik}A_{kj} -\frac{\p^ 2 p}{\p x_i \p x_j} + \nu \frac{\p^2 A_{ij}}{\p x_m \p x_m}, 
\ee
where $d/dt$ is the convective derivative, $p$ stands for pressure divided by fluid density and $\nu$ corresponds to the kinematical viscosity. In equation (\ref{1}), $A_{ij} = \partial_j u_i$ is the velocity gradient tensor in cartesian components. The difficulty in obtaining statistics from the velocity gradient Navier-Stokes is that the pressure Hessian and the viscous term are not closed in terms of a Lagrangian trajectory.  A review of different attempts of closures can be found at \cite{meneveau}. The simplest closure is achieved by neglecting dissipation and nonlocal effects of the pressure Hessian. Although, a solution is available, it can be shown that it develops a divergence at finite time \cite{viei}, \cite{cant}. The RFD has the merit of incorporating pressure and viscous effects preventing divergences in $\A$.  It may be compared to the tetrad model \cite{chertkov_etal}, though instead of dealing with an equation for the  evolution of fluid deformation, it is strongly modelled. The rationale goes as follows. 
Write the pressure Hessian  as
\be
\lb{1a}
\frac{\p^ 2 p}{\p x_i \p x_j} \approx  \frac{\p X_m}{\p x_i}\, \frac{\p X_n}{\p x_j}\; \frac{\p^ 2 p}{\p X_m \p X_n} 
\ee
where ${\p X_j}/{\p x_i}$ denotes the Jacobian of the change of coordinates from Eulerian to Lagrangian coordinates. In (\ref{1a}), spatial derivatives of the Jacobian were neglected.
The Cauchy-Green tensor, defined by
\be
\lb{2a}
C_{ij}=  \frac{\p x_i}{\p X_k}  \frac{\p x_j}{\p X_k} 
\ee	
is assumed to have the form
\be
\lb{3a}
\C = \exp [ \tau \A] \exp [ \tau \AT ] \ ,
\ee
where $\tau$ corresponds to a short time associated to the correlation time of the velocity gradient in the Lagrangian frame, assumed to be of the order of the Kolmogorov time scale.    
The idea behind the RFD approximation is that after a short period of time ($\sim \tau$) the shape of a Lagrangian particle is uncorrelated with its initial shape. Therefore, it is possible to assume an isotropic shape for a fluid particle at initial time, which implies an isotropic pressure Hessian $\frac{\p^ 2 p}{\p X_m \p X_n} = \frac 1 3 \delta_{mn}\frac{\p^2 p}{\p X_l \p X_l}$. Taking it into account, (\ref{1a}) turns to 
\be
\lb{2}
\frac{\p^ 2 p}{\p x_i \p x_j} \approx \frac{C^{-1}_{ij}}{C^ {-1}_{qq}} A_{mn}A_{nm}.
\ee
Similar reasoning can be applied to model the viscous term, yielding
\be
\lb{3}
\nu \frac{\p^ 2 A_{ij}}{\p x_m \p x_m}\approx  \frac{\p X_k}{\p x_m} \frac{\p X_l}{\p x_m} \frac{\p^ 2 A_{ij}}{\p X_k \p X_l} \approx - \frac{1}{3T} C^{-1}_{qq}A_{ij} 
\ee
where $T$ stands for the integral time scale, which comes from dimensional arguments as $\nu/ (\p X)^2 \approx 1/T$, considering that $\p X$ is on the order of a typical distance travelled by a particle during time $\tau$, which scales with the Taylor microscale length.
Therefore, substituting eqs. (\ref{2}) and (\ref{3}) in (\ref{1}), the RFD model equation is given by
\be
\lb{4}
\dot \A =  - \A^2 +  \frac{\C^{-1} \Tr( \A^2) }{\Tr(\C^{-1})} - \frac{\Tr(\C^{-1})}{3T} \A + g \mathbb{F} \, ,
\ee
where a random forcing was supplemented to provide stationary statistics. In (\ref{4}), $g$ is the strength of the stochastic force, related to energy injection rate, and will play an important role in the discussion.  $\F$ is a zero average white noise tensor such that
\be
\lb{5}
\langle F_{ij}(t) F_{kl}(t') \rangle = G_{ijkl}  \delta(t-t') \ ,
\ee
with
\be
\lb{6}
G_{ijkl}=2 \delta_{ik} \delta_{jl} - \frac{1}{2} \delta_{il} \delta_{jk}- \frac{1}{2} \delta_{ij} \delta_{kl} \ . \
\ee
The force correlator $G_{ijkl}$ is the general $4^{\text{th}}$-order tensor which respects isotropy and also ensures incompressibility, \emph{i.e.}, $\Tr \,\A = 0$. It can be shown that    $G_{jjkl} = 0$  and $G_{ijkl}= G_{klij}$, which follow immediately from equation (\ref{5}). 

\section{Results}

\subsection{Instantons in the Martin-Siggia-Rose path integral}

As in many applications of large deviations, it is customary to evaluate the  probability to reach a final state $\A (t_2) = \A_2$ at time $t=t_2$ starting from time $t_1$, with $\A (t_1) = \A_1$. The initial configuration $\A_1$ is usually taken to be at, or close to, an attractor of the deterministic dynamics, whilst the initial time is assumed to be $t_1 = -\infty$, such that the stationary transition probability will depend solely on $\A(t_2)$.  In this work,  we want to evaluate the probability of finding a large value of one component $A_{\alpha \beta}(t_2)$, either longitudinal, or transverse, which can be  accomplished with the auxiliary of the Martin-Siggia-Rose/Janseen/de Dominics  path integral  \cite{msr}, \cite{janssen}, \cite{dominicis}. Therefore, denoting the referred  transition probability by $\rho_{\alpha \beta}(a) = \rho(A_{\alpha \beta}(t_2) = a |\A(t_1) = 0) $ with $\alpha$ and $\beta$ prescribed ( $\rho_{\alpha \beta}$ should not be understood as a tensor, the indices simply refer to the transition probability  of the component $A_{\alpha \beta}$ to the value $a$ at a final time $t_2$), the path integral formalism leads to
\be
\lb{A7}
\rho_{\alpha \beta}(a) = \langle \delta(A_{\alpha \beta}(t_2) - a) \rangle = \int \, D[\A]  \exp\left[- \int_{t_1} ^{t_2} dt \, \Lag_{\text{OM}} [\A(t),\dot \A(t)] \right] \,  \delta(A_{\alpha \beta}(t_2) - a) \, ,
\ee
where the angular brackets stand for the averaging over force realisations, which can be accounted for, alternatively, by performing a sum over all possible paths $\A(t)$ starting from $\A_1$ and arriving at $\A_2$. The final condition is enforced by the Dirac delta functional, and the Onsager-Machlup Lagrangian $\Lag_{\text{OM}}[\A(t),\dot \A(t)]$ \cite{onsager-machlup} reads
\be
\lb{A4}
\Lag_{\text{OM}}[\A,\dot \A] =\frac{1}{2g^2}\left(  (\dot A_{ij} - V_{ij}) Q^{-1}_{ijkl}(\dot A_{kl} - V_{kl})  - \frac{1}{5} \text{Tr}[\dot \A - \V]^2 \right),
\ee
with $Q^{-1}_{ijkl}= (8/15) \delta_{ik}\delta_{jl} +(-2/15)\delta_{il}\delta_{jk}$ such that $G_{ijkl} = Q_{ijkl} - Q_{ijmm}Q_{klnn}/Q_{ppqq}$.
Equivalently, the probability transition can be written in terms of the Martin-Siggia-Rose Lagrangian $\Lag_{\text{MSR}} [\A(t),\hat \A(t)] $ as
\be
\lb{A6}
\rho_{\alpha \beta}(a) = \langle \delta(A_{\alpha \beta}(t_2) - a) \rangle = \int \, D[\A] D[\hat \A] \exp\left[- \int_{t_1} ^{t_2} dt \, \Lag_{\text{MSR}} [\A(t),\hat \A(t)] \right] \,  \delta(A_{\alpha \beta}(t_2) - a) \, ,
\ee
with
\be
\lb{A2}
\Lag_{\text{MSR}}[\A,\hat \A] =  \frac{g^2}{2} \hat A _{ij} G_{ijkl} \hat A_{kl} - i \,\text{Tr} [\hAT (\dot \A - \V )]  \, .
\ee
The relationship between the two Lagrangians is made clearer by noting that from (\ref{A2}) the conjugated momentum reads $\Pe =\p \Lag / \p \dot \A = -i \,\hat \A$, so the Lagrangians are related by a Legendre transform. In order to obtain the instanton equations we have, thus, to derive the stationary action (\ref{A2}) with the endpoint $A_{\alpha \beta}(0)  = a$ imposed by the Dirac delta functional.  We are going to clarify how this constraint turns to a final condition for the auxiliary variable $\Pe(t)$, since this point is not usually discussed in the literature. Many authors consider the physical reasoning of Guraire and Migdal \cite{migdal} based on the negative viscosity sign. The limitation of this argument is that it applies only to fluid systems. The discussion below encompasses more general cases.

Starting from the Onsager-Machlup Lagrangian we calculate  the action variation with respect to the path $\A(t)$ with initial point fixed, that is, $\delta \A(t_1) = 0$, yielding
\begin{align}
\lb{5ap}
\delta S & = \int_{t_1} ^{t_2} dt \; \left\{\text{Tr}\left [  \frac{\p \Lag}{\p \A} \delta \A^{T}(t) + \frac{\p \Lag}{\p \dot \A} \delta \dot \A^{T}(t) \right ]  + \lambda \delta(t-t_2)\,\delta A_{\alpha \beta}(t)]  \right\}\\
\lb{6ap}
& = \int_{t_1} ^{t_2} dt \; \text{Tr}\left [  \frac{\p \Lag}{\p \A} \delta \A^{T}(t) + \frac{d}{dt}\left( \frac{\p \Lag}{\p \dot \A} \delta \A^{T}(t) \right) -  \frac{d}{dt}\frac{\p \Lag}{\p \dot \A} \delta \A^{T}(t) \right] + \nn \\ 
& \hspace{2cm}+ \lim_{\epsilon \rightarrow 0} \int_{t_1} ^{t_2+\epsilon}dt \;  \lambda \, \delta(t-t_2) \,\delta A_{\alpha \beta}(t) \\
\lb{7ap}
& = \int_{t_1} ^{t_2} dt \;\text{Tr} \left [ \left( \frac{\p \Lag}{\p \A} -\frac{d}{dt}\frac{\p \Lag}{\p \dot \A} \right)\delta \A^{T}(t) \right]  + \text{Tr} \left( \frac{\p \Lag}{\p \dot \A} \delta \A^{T}(t) \right) _{t_1} ^{t_2} +\lambda \, \delta A_{\alpha \beta}(t_2)]
\end{align}
The last term in (\ref{5ap}) is due to writing the Dirac delta in terms of its Fourier representation. By demanding the action variation to be stationary with respect to the path $\A(t)$ we arrive at
\begin{align}
\lb{8ap}
 & \frac{\p \Lag}{\p \A} -\frac{d}{dt}\frac{\p \Lag}{\p \dot \A} = 0 \\
 & \text{Tr}[\Pe(t_2)\, \delta \A^{T} (t_2)] - \text{Tr} [\Pe(t_1)\, \underbrace{\delta \A^{T}(t_1)]}_{= 0} + \lambda \, \delta A_{\alpha \beta}(t_2)]=0 \lb{9ap},
 \end{align}
where we used the definition $\Pe(t) \equiv \p \Lag /\p \dot \A(t) $. Equation (\ref{8ap}) is the Euler-Lagrange equation which gives the evolution with time, while (\ref{9ap}) implies $P_{ij}(t_2)  = -\delta_{i\alpha}\delta_{j\beta}\lambda$. This completes our derivation relating the final point condition of $\A$ with $\Pe(t)$. Note that in this case, the endpoint is not fixed as usual.  The Dirac delta relaxed the endpoint, allowing it to  have non vanishing variation  ($\delta \, \A(t_2) \neq 0$). 

Therefore, since a final condition for the canonical momentum is obtained, it is more convenient to minimize the MSR action rather than minimising the OM Lagrangian, since the former is first order in time. Hence, substituting $-i \hat \A(t)$ by $\Pe(t)$ in (\ref{A2}), we are led to solve the set of saddle-point equations 
\ben
\lb{11}
&\dot A_{ij} = V_{ij}(\A) + g ^2 G_{ijkl} P _{kl}  \\
\lb{12}
&\dot P _{ij} = -P _{kl} \nabla_{ij}V_{kl}(\A) \, , 
\een
with endpoint condition $P_{ij}(t_2)  = -\lambda \, \delta_{i\alpha}\delta_{j\beta}$. The solution of equations (\ref{11}) and (\ref{12})  minimize the MSR action (\ref{A2}), (\ref{A6}) subject to the endpoint constraint $A_{\alpha \beta}(t_2) = a$. 

Thus, we end up with a system of mixed initial-final condition which naturally suggests that $\Pe$ should be integrated backwards in time whereas $\A$ is integrated forwards in time. 
This kind of problem was tackled numerically by Chernyk and Stepanov \cite{stepanov} and by \cite{grafke_schafer}, \cite{grafke_eric} in the context of the Burgers equation. There they were seeking large values of the velocity gradient in one point. It was found that the instantons turned out to be the shocks which are present in the underlying dynamics of the system. See also \cite{instanton_num_review} for a review of applications of this approach, including the study of  instantons in the stochastic Navier-Stokes equation.

By scaling the auxiliary variable $ \tilde \Pe = g^2 \Pe$  the action  changes as $\mathcal S [\Pe, \A] \rightarrow \tilde {\mathcal S} [\tilde \Pe, \A]/g^2$, yielding for the conditional probability distribution
\be
\lb{10a}
\rho_{\alpha \beta}(a) = \int  D[\Pe] D[\A] \, \delta[A_{\alpha \beta}(0)  - a]  \exp \left \{ - \frac{\tilde {\mathcal S}[\tilde \Pe, \A]}{g^2}  \right \} 
\ee
where $\tilde {\mathcal S}[\tilde \Pe, \A]$ is independent of $g$.  In the weak noise limit $g\rightarrow 0$, the probability $\rho_{\alpha \beta}(a)$ will be dominated by the contribution from the action minimizer. This is in accordance with the Freidlin-Wentzell theory of large deviations \cite{freidlin_wentzel}, which states that
\be
\lb{11a}
- \lim_{g \to 0} \; g^2 \ln \rho_{\alpha \beta}(a) = \mathcal I = \text{min} \, \tilde{\mathcal S}
\ee
 where the action minima $\text{min} \, \tilde{\mathcal S}$, is evaluated at the optimal path satisfying $A_{\alpha \beta}(t_2=0)=a$ and  $\A(t_1=-\infty)=0$. The rate function $\mathcal{I}$, independent of $g$, controls the behavior of the transition probability for asymptotically vanishing $g$. It contains information not just about small fluctuations around the attractor of $\A$ ($\A =0$ for the dynamics considered) but also about large fluctuations.



For the sake of clarity, we split the cases where the fixed final value of the velocity gradient is either one of the diagonal (longitudinal) or off-diagonal (transverse) components. 

\subsection{Transverse gradient statistics}
This section shows the results regarding the stationary statistics $\rho_{12}(a)$ of the transverse velocity gradient. Due to numerical reasons we should use a finite but large initial time $t_1$. In our implementation we chose $t_1= -6 T$ whereas $t_2=0$, that is, the evolution is carried out through six integral time scales. It was also checked numerically that this value suffices for stationarity by examining  the time series of the original SDE, integrated according to \cite{kloeden-platen}.  The algorithm is an iterative procedure to obtain the solution of the set of equations (\ref{11}) and (\ref{12}). 

Before we apply the method treating the eight independent degrees of freedom encoded in $\A$, it is convenient to take advantage of the symmetries of the problem in order to reduce the number of degrees of freedom, lowering thus the computational cost. First, we recall the Onsager-Machlup action (\ref{A7}). After we write the Dirac delta using its Fourier representation, the endpoint condition can be understood as another term in the action of the form $\lambda \delta(t-t_2)\, A_{\alpha \beta}(t)$ (in this section $(\alpha,\beta)  = (1,2)$). This additional term, which manifests itself in the equations of motion (\ref{11}) and (\ref{12}) as a final condition for $\Pe$, breaks the parity symmetry  $x_i \rightarrow -x_i$ and $v_i \rightarrow -v_i$ for $i = 1,2$, therefore only the symmetry $x_3 \rightarrow -x_3$ and $v_3 \rightarrow -v_3$ remains. If the action exhibits this symmetry so does the solution to the equations of motion, provided the final/initial conditions keep the same symmetry, which is the case. Hence $\A$ must be a velocity gradient tensor with reflection symmetry in the $x_3$  direction,  whose only possible form is
\be
\lb{transversal}
\A(t)= \left(
\begin{array}{ccc}
 A_{11}(t)  &  A_{12}(t)  &  0   \\
 A_{21}(t)  &  A_{22}(t)  &  0   \\
 0          &      0      &  -A_{11}(t)-A_{22}(t)
\end{array}
\right)\, .
\ee
We are left with 4 independent variables instead of eight, which simplifies the computation considerably.  

\begin{figure}
\centering
\hskip25pt\input{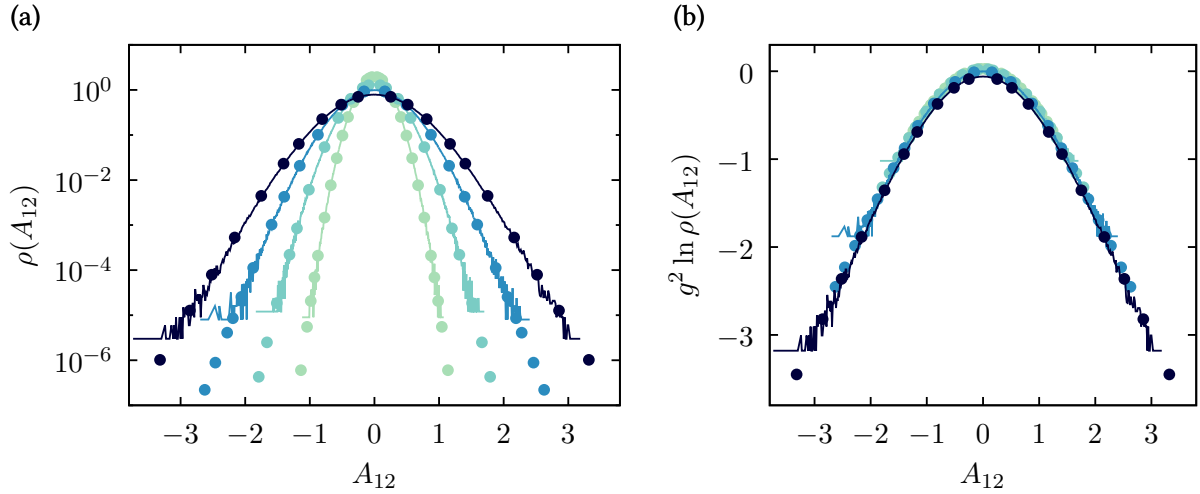}
\caption{(a) Semilog plot of the transverse velocity gradient pdf. Dots: numerical instanton evaluation.  Solid lines: pdfs from SDE (\ref{4}). The range of $A_{12}$ lies between 5 to 6 standard deviations. Forcing values are $g=0.2 \, , 0.3 \, , 0.4 $ and $0.5$ where darker colours correspond to higher values of $g$. (b) Rescaled pdfs corresponding to vertical axis $g^2 \ln \rho(a)$ showing collapse.}
\label{fig1}
\end{figure}

Now, the Chernyk-Stepanov method can be performed. The idea is to decouple $\Pe(t)$ and $\A(t)$ for the first iteration. For instance, we set $\A(t) =0$ and solve (\ref{12}) backwards in time for an arbitrarily chosen $\lambda$. In the next step, we substitute the time series of $\Pe(t)$ obtained in (\ref{11}), which is integrated forward in time to obtain $\A(t)$. This is performed recursively until the solutions converge. Both equations are solved by the 4th order Runge-Kutta scheme with time step $dt=10^{-3}$ and a piecewise cubic interpolation is performed to obtain the intermediate time steps required by the method. The criteria used for convergence is that $|A_{12}(0) - A_{12}^{\text{old}}(0)|/|A_{12}^{\text{old}}(0)| < \delta  $, \emph{i.e}, the relative error of the obtained instanton in comparison with the (old) instanton calculated in the previous iteration should be smaller than a quantity $\delta$ (we set $\delta=10^{-7}$ and $\delta = 10^{-10}$ for the longitudinal case).  With the instanton solution, the probability of arriving at a final value $A_{12} = a$ can be computed plugging it into (\ref{11a}). Spanning a set o $\lambda$'s we can generate the pdf $\rho_{12}(a)$, since each value of $\lambda$ leads to a different final value of the longitudinal velocity gradient $a$. Figure  \ref{fig1}(a) displays  pdfs obtained by this approach for different values of forcing amplitude. Pdfs from numerical integration of the SDE are also plotted for comparison, showing good agreement between the results.
The collapse depicted in figure \ref{fig1}(b) corresponds to a rescaling of the vertical axis, $g^2 \ln (\rho(A_{12}(t_2)=a))$ and shows that the pdfs calculated obey the large deviation principle (\ref{11a}). The curve is minus the rate function (action minima) as a function of the final value $A_{12}$.

A last comment on the numerical scheme  concerns  convergence issues that may arise. Actually, in the original reference of the method  \cite{stepanov} it was reported that, for a critical value of $\lambda$, the numerical convergence becomes problematic. In our case, it is noticed that as $| \lambda |$ increases, so does the number of iterations to reach convergence. In the transverse case, where the number of degrees of freedom cannot  be as reduced as in the longitudinal case (cf. next subsection), convergence may fail completely. In order to circumvent this issue we performed the following strategy. Let $A^{\alpha}$ and $P^{\alpha}$  be the $\alpha$-th step in the iteration procedure of the numerical integration. The direct approach would be to use the series $A^{\alpha}$ and $P^{\alpha}$ in the saddle-point equation (\ref{12}) to obtain $P^{\alpha+1}$ and $A^{\alpha+1}$ and so on. However, when the iteration ceases to converge, we modify $A^{\alpha+1}$ by $A^{\alpha + 1} \rightarrow \beta A^{\alpha}+ (1-\beta)A^{\alpha+1}$, with $\beta$ arbitrarily chosen on the interval $[0,1]$, that is, the next iteration is a weighted average of the old and the new ones. Although not systematic, since we do not know \emph{a priori} which is the optimal $\beta$ value, this procedure dumps large variations in each step and tends to keep iterations inside the converge radii. Values as big as $\beta = 0.8$ may be needed to capture the tail of the distributions.

\subsection{Longitudinal gradient statistics }
In this section we show the results concerning the longitudinal velocity gradient. In order to calculate the instanton we make use of the even higher degree of symmetry of this case, which reduces the number of degrees of freedom to only one. We invoke the same rationale of the previous section. The difference is that imposing $A_{11}(0) =a$ consequently adds to the action a term that respects parity symmetry $x_i \rightarrow -x_i$, $v_i \rightarrow - v_i$ in all directions and hence implies that the instanton velocity gradient must be diagonal. This term breaks rotation symmetry though, by selecting the $x_1$ direction, but the action is still invariant under rotations around the $x_1$ axis. So, the action makes no preference between the $x_2$ or $x_3$ directions, implying $A_{22} = A_{33}$ for the solution.  Moreover, incompressibility leads to $\A =\text{diag} (A(t),-A(t)/2,-A(t)/2)$, {\it i.e}, the velocity gradient depends on a single degree of freedom.  Within this simplification the saddle-point equations become much faster and stable to be integrated numerically.

\begin{figure}
\centering
\hskip25pt\input{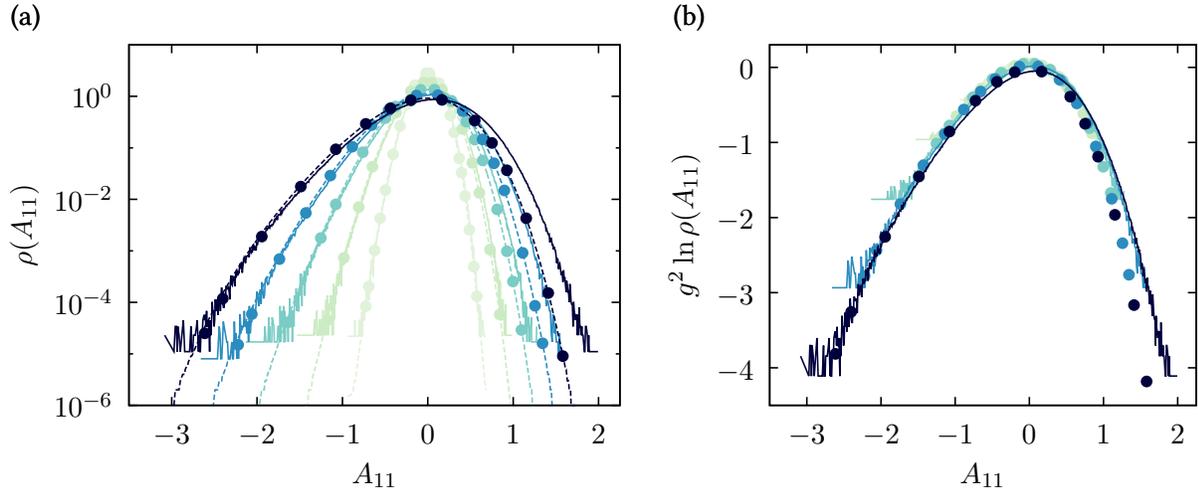}
\caption{(a) Semilog plot of the longitudinal velocity gradient pdf. Dots: numerical instanton evaluation.  Solid lines: pdfs from SDE (\ref{4}). Dashed lines: analytical (\ref{21}). The range of $A_{11}$ lies between 5 to 6 standard deviations. Forcing values are $g=0.2 \, , 0.3 \, , 0.4 \, , 0.5$ and $0.6$ where darker plots correspond to higher $g$ values. (b) Rescaled pdfs corresponding to vertical axis $g^2 \ln \rho(a)$ showing collapse.}
\label{fig2}
\end{figure}

Apart from the numerical solution to the saddle-point equations, the high degree of symmetry enables us to derive an analytical solution in the case of longitudinal velocity gradient.  With the velocity gradient given by a diagonal form $ \A =\text{diag} (A(t),-A(t)/2,-A(t)/2)$, a reduced MSR action for the single degree of freedom $A(t)$ can be written as
\be
\lb{18}
S_{\text{red}} [A,p] = \int _{-T}^{0} dt \, \left [  p \left( \dot A - b[A] \right) - \frac{g^2}{2} p^2 \right] \, ,
\ee
where $A(t)$ is a scalar, equivalent to the $A_{11}$ of the original system and $b[A] = V_{11}[\A]$. Due to this drastic reduction of degrees of freedom, it is possible to write  $b[A]$ as a gradient of a function $h[a]$
\be
\lb{19}
b[A] = -\nabla h[A], \quad h[A] = \frac{A^2}{2}  + \frac{A^3}{6} +\frac{ \tau}{4}(1 + \tau)A^4 - \frac{\tau^2}{10}A^5 + \mathcal O (\tau^3) \, .
\ee
In that case,  instantons may be obtained as the reverse of the relaxation path from $A(0)$ to $A(-\infty)$ \cite{bouchet_etal}. Nevertheless, the pdf can be computed in a more straightforward manner by solving the corresponding Fokker-Planck equation. First, we write an effective SDE which leads to the above reduced action (\ref{18})
\be
\lb{20}
\dot A = b[A] + g f(t) \, ,
\ee
where $\langle f(t) f(t') \rangle = \delta(t-t')$ is the correlation of the reduced noise $f(t)$. A straightforward calculation shows that the MSR action related to the SDE (\ref{20}) is given by (\ref{18}). The Fokker-Planck equation can be easily derived from (\ref{20}), whose stationary solution reads
\be
\lb{21}
\rho(a) = N \,  \exp(-2h[a]/g^2) \, ,
\ee
with $h[A]$ given by (\ref{19}) and $N$ is normalization factor. This important result validates the numerical procedure, as one can see in figure \ref{fig2}(a), where a good agreement between the analytical and numerical instanton contribution to the pdf is achieved.

Once the pdf $\rho(a)$ is obtained analytically, it is possible to evaluate the moments of the velocity gradient as a power series of the noise $g$ along with the scaling with Reynolds number, which is another original result of this paper.  A straightforward computation yields for the first central moments of the longitudinal velocity gradient,
\begin{align}
&\text{var}[a] = \frac{g^2}{2} + \frac{g^4}{96} (29 - 180 \tau(1+\tau)) \, , \label{variance}\\
&\frac{\E[(a-\E[a])^3]}{\text{var}^{3/2}[a]} = -\frac{g}{\sqrt{2}} + 
 g^3 \left(-\frac{25}{24 \sqrt{2}} + \frac{15\,\tau}{\sqrt{2}} + 9 \sqrt{2} \tau^2 \right)  \, ,\label{skewness}\\
& \frac{\E[(a-\E[a])^4]}{\text{var}^{2}[a]} =  3 + \frac{1}{16} g^2 (19 - 60\, \tau(1+\tau)) \, .\label{flatness}
\end{align}
We highlight this is a novel result specially considering there are few analytical results concerning velocity gradient models available.  
Let us compare it to phenomenological expectations.  The forcing $g$ may be interpreted as the energy injection rate in the Lagrangian particle per unit area. Since stationarity demands that energy injection equals  energy dissipation, the stochastic RFD equation (\ref{4}) leads to $g^2 \sim  \p^2 \varepsilon/(\p x)^2 \sim \ep /\lambda^2$, where $\varepsilon $ is the dissipation rate and $\lambda$ is the Taylor microscale length. On dimensional grounds one would expect the velocity gradient variance to behave as $\langle (\p u)^2 \rangle \sim \ep /\nu \sim \ep \, Re/(UL)$ ($U$ is a typical integral velocity scale) which implies $\langle (\p u)^2 \rangle \sim g^2 \lambda^2 \, Re/(UL) \sim g^2 T$ in agreement with (\ref{variance}) at least to leading order (recall we have set $T=1$). 

\begin{figure}
\centering
\hskip20pt\input{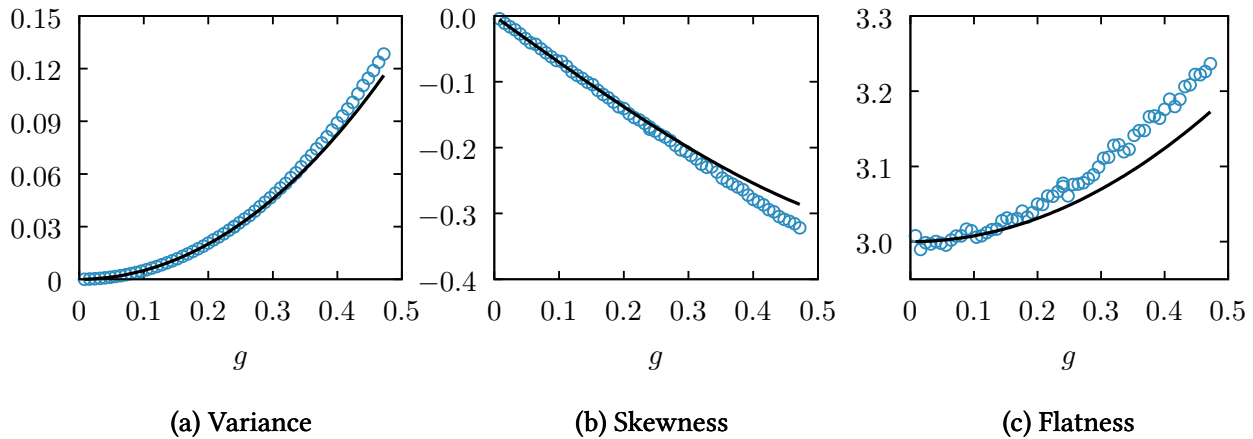}
\caption{Statistical moments of the longitudinal velocity gradient as a function of the forcing $g$. Circles: numerical integration. Solid lines: instanton analytical results (equations \ref{variance}-\ref{flatness}).}
\label{fig3}
\end{figure}

Comparison with the numerical solution of the SDE, figure \ref{fig3}, shows compatibility between analytical and numerical moments for small values of forcing.  As $g$ increases though, the analytical result disagrees with the numerical evaluation since for finite $g$ the instanton approximation is not sufficient to estimate the pdf. Moreover, it can be also noted that the agreement between numerical and  analytical moments decreases for higher moments, which is expected considering the analytical pdfs mismatch the numerical ones in the tails (specially the right tail), figure \ref{fig2}.



 The skewness and flatness, though, show an incorrect scaling with respect to Reynolds number which points to a drawback of the model.  
This drawback appearing at high Reynolds numbers was already recognized in Ref. \cite{CheMen07}. The new analytical results provided in Eqs. (\ref{variance}), (\ref{skewness}), (\ref{flatness}) shed a new light on the numerical results obtained in this reference  \cite{CheMen07}. Indeed, it is there underlined that the variance of the gradients does not behave with the free parameter of the model, i.e. the Reynolds number, in a consistent way with the dimensional approach of Kolmogorov. To circumvent this issue, it was proposed instead in Ref. \cite{CheMen07} to study the relative scaling of the logarithm of higher order moments of the gradients with respect to the variance of the gradients. To interpret the departure of the observed scalings seen in Ref. \cite{CheMen07} from non intermittent scalings, it is then tempting to interpret them, based on the theoretical results of Eqs. (\ref{variance}), (\ref{skewness}), (\ref{flatness}), as being reminiscent of the forcing. Future works will be devoted to improve the RFD approximation in order to include genuine intermittent scalings, at the cost, perhaps, of introducing a further free parameter that quantifies in an appropriate way intermittent corrections. We leave these perspectives for future investigations.

Regarding the so-called $RQ$ plane, the velocity gradient instanton starts at $\A_1 =0$ evolving to a final configuration such that $A_{11}(0) =a$. If we keep track of the trajectory on the $RQ$ plane it is  noticed that it lies entirely in the Vieillefosse line ($4 Q^3 + 27R^2 = 0$, with $Q = - \text{Tr}\, \A^2 /2$ and $R =   - \text{Tr} \, \A^3 /3$) \cite{viei}, although this is not a consequence of the model dynamics. Actually, this is simply due to kinematics since for a velocity gradient tensor taking the form $\A =\text{diag} (A(t),-A(t)/2,-A(t)/2)$, which in turn is a consequence of symmetry, the Vieillefosse line is satisfied identically.

\subsection{Filtering and interpretation of instanton solution}
\begin{figure}
\centering
\hskip40pt\input{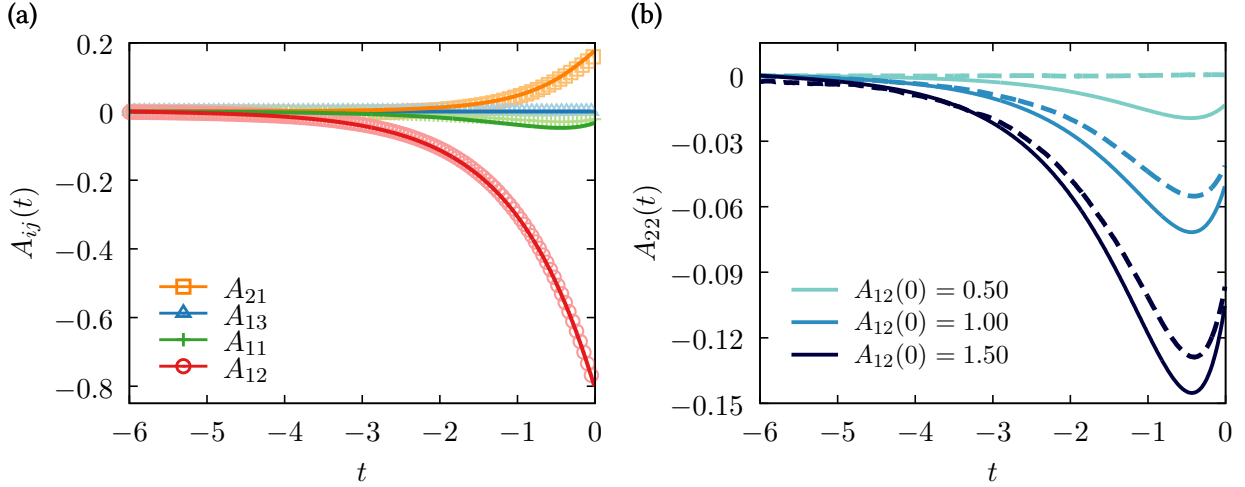}
\caption{(a) Different components from conditionally averaged trajectories (symbols) compared to the instanton (solid) for the case where $A_{12}$ is set to $-0.8$ at the endpoint. Components $A_{13}$, $A_{23}$, $A_{31}$, $A_{32}$ (not all shown) are negligible, in agreement with our symmetry argument. (b) Component $A_{22}$ both from filtering (dashed) and instanton (solid) for the case where $A_{12}$ is set to $0.5$, $1.0$ and $1.5$ at the endpoint. In both figures $g=0.5$.}
\label{filtered_offdiag}
\end{figure}

In this subsection we try to assess the relevance of instantons in a fluid dynamical model sharing many non trivial properties with real turbulence, as it is the case for the RFD approximation, following reference \cite{grafke_schafer}. An ensemble with trajectories of the original SDE without any constraint was build. With this ensemble we perform a conditioned statistics selecting those paths ending within a small neighborhood of $a$, that is, $A_{11}(0) \in [a-da,a+da]$ ($A_{12}(0) \in [a-da,a+da]$ if we are looking at transverse gradients).  To increase the ensemble sizes, if the searched value $a$ is crossed by any component, we perform frame rotations over the entire trajectory so that it always corresponds to component $A_{11}$ (in the diagonal case) or $A_{12}$ (off-diagonal). What is seen is that these paths concentrates around the instanton solution and after being averaged they tend to superpose with it as depicted in figure \ref{filtered_offdiag}. Figure \ref{filtered_offdiag}(a) shows several components of velocity gradients from conditionally averaged trajectories compared to the instanton solution with final value $A_{12}(0) = -0.8$. Figure \ref{filtered_offdiag}(b) depicts how the unconditioned component $A_{22}(t)$  evolves for different final values of the conditioned $A_{12}(0)$ in comparison with instanton solution.  The agreement is better as the constrained final value gets larger, as expected by instanton theory. This trend has been found in the context of Burgers equation in \cite{grafke_schafer} and \cite{grafke_eric}.  After all it is clearly obtained that typical trajectories of the stochastic dynamics fluctuates around but not far from the instanton trajectory provided $g$ is small in accordance with the large deviation principle.

\begin{figure}
\centering
\hskip40pt\input{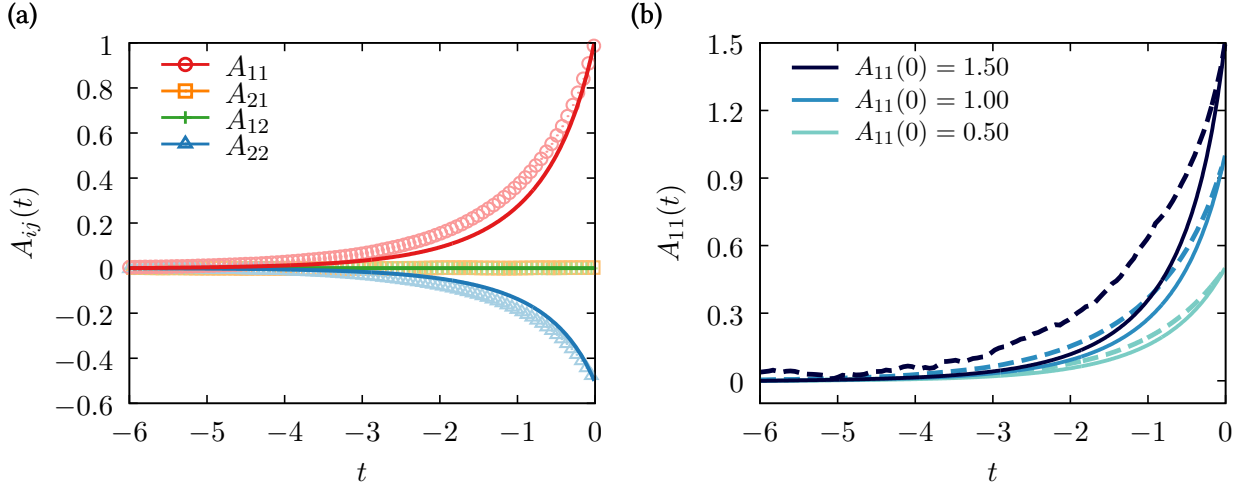}
  \caption{(a) Different components from conditionally averaged trajectories (symbols) compared to the instanton (solid) for the case where $A_{11}$ is set to $1.0$ at the endpoint.  Components $A_{21}$, $A_{12}$ are negligible (as well as other off-diagonal components not shown), in agreement with our symmetry argument. (b) Component $A_{11}$ both from filtering (dashed) and instanton (solid) for the case where $A_{11}$ is set to $0.5$, $1.0$ and $1.5$ at the endpoint. In both figures $g=0.5$.}
   \label{filtered_diag}
\end{figure}

Conversely, the most probable trajectory leading to a certain value of longitudinal velocity gradient is such that the velocity gradient is diagonal, as claimed in section IIIB by symmetry arguments. This statement is indeed confirmed by the filtering procedure as presented in figure \ref{filtered_diag}. Figure \ref{filtered_diag}(a) shows the average behaviour of velocity gradient conditioned to $A_{11}(0) = 1.0$ in comparison with instanton trajectories.  All off-diagonal components vanish, as illustrated by $A_{12}(t)$ and $A_{21}(t)$  (others not shown). In figure \ref{filtered_diag}(b) three different constrained values are exhibited. In contrast to the previous case the agreement does not improve for larger values of $A_{11}(0)$, another manifestation of the mismatch observed on the tails of the diagonal pdf (figure \ref{fig2}).

\section{Conclusion}
The role of the rare events can be revealed by means of the Martin-Siggia-Rose path integral formulation. In this work we apply this technique to a model of Lagrangian turbulence called the Recent Fluid Deformation (RFD). This closure comprises a stochastic model of the velocity gradient based on short time correlations in the Lagrangian frame. Within the path integral formalism the most probable trajectory that leads to a certain event is calculated numerically and,  for the longitudinal velocity gradient case, also analytically.  We showed the use of symmetries can rule out unnecessary degrees of freedom allowing  less numerical effort in order to compute the instanton. Apart from the benefited numerical computation, the symmetries let us evaluate an analytical approximated solution for the longitudinal velocity gradient pdf, enabling us to unveil its central moments dependence on the Reynolds number. 


Both longitudinal and transverse cases present the instanton satisfying the Vielleifosse line. We believe that the rationale for that lies in the dominance of the non deviatoric terms figuring the model equation (\ref{4}). That is, when $\tau \rightarrow 0$, the RFD approximation approaches the Restricted Euler equation.  

Regarding vorticity alignment, instanton solutions for transverse gradients shows a complete alignement with the intermediate strain eigenvalue, which can be seen computing the normalized product of the three rate of strain eigenvalues  $s^* = -3\sqrt{6} \lambda_1 \lambda_2 \lambda_3/(\lambda_1^2+ \lambda_2^2+ \lambda_3^2)^{3/2}$ \cite{lund_rogers}, resulting $s^* = 1$, where $\lambda_i$, $i = 1, \,2, \, 3$, are the referred eigenvalues. Since the instanton corresponds to the most probable trajectory leading to a certain value of the velocity gradient, our result agrees with reference \cite{lund_rogers} which showed that the pdf of $s^*$ develops a sharp peak around $s^* = 1$.

The longitudinal velocity gradient pdf has a weaker agreement in comparison with the transverse one as the forcing increases, showing the instanton approximation is not enough to account for the full statistics even for moderately low values of $g$. It means that fluctuations around the instanton solution may play an essential role, which could be hopefully analyzed by perturbative methods. Perturbative corrections to the instanton pdf can be dealt with the effective action approach \cite{effectiveaction} and is currently under study. The issue of wether the instanton approach suffices and perturbative methods are fit to more complex fluid dynamical systems is an important matter and deserves further investigation.

As a final remark, the application of the instanton study to this Lagrangian model allowed us to understand the scaling of the statistical moments with the Reynolds numbers. This opens new possibilities in the direction of refinement of the RFD approximation in order to grasp more aspects of the phenomenology of turbulence. Moreover we expect that the use of symmetries as in this work, which led to a reduction of the degrees of freedom, can be applied to other stochastic systems allowing more efficient optimal paths computation.

\section{Acknowledgments}
L.S.G. would like to thank the financial support by CAPES scholarship program, process number $99999.006843/2014-00$ and also the warm hospitality of the Laboratoire de Physique, \'{E}cole Normale Sup\'{e}rieure de Lyon where this work was developed. R.M.P. thanks CAPES for financial support through the scholarship process number 9497/13-7.

\end{document}